# Towards Developing an Understanding of Consumers' Perceived Privacy Violations in Online Advertising

**Kinshuk Jerath**, Arthur F. Burns Professor of Free and Competitive Enterprise, Chair of the Marketing Division, Columbia University, USA, kj2323@columbia.edu

**Klaus M. Miller**, Assistant Professor, HEC Paris, Chairholder at the Hi!PARIS Center on Data Analytics and Artificial Intelligence for Science, Business and Society, France, millerk@hec.fr

**D. Daniel Sokol**, Carolyn Craig Franklin Chair in Law and a Professor of Law and Business, University of Southern California, USA, dsokol@usc.edu

Acknowledgments: We thank Alessandro Acquisti, Laura Brandimarte, Avi Goldfarb, Kathrin Gruber, Garrett Johnson, Tesary Lin, and Timo Mueller-Tribbensee for their helpful comments on an earlier draft of this paper. This manuscript has also benefited from comments from seminar and conference participants at Columbia University, Paris Dauphine University, the University of Chile, the FTC PrivacyCon 2024, BIOMS 2024 at the University of Michigan, the S&O Research Day 2024 at HEC Paris, the CNIL Privacy Research Day 2024, SCECR 2024, ZEW ICT Conference 2024, YES Marketing 2024, and the Mapping and Governing the Online World Workshop at NYU Law School 2025.

**Towards Developing an Understanding of Consumers' Perceived Privacy Violations in Online Advertising**

**Abstract**

Privacy-enhancing technologies (PETs) represent a critical operational challenge for the online advertising industry, requiring substantial infrastructure investment while promising improved consumer privacy protection. Even when PETs may improve privacy protection from an operational or technical viewpoint, understanding whether PETs actually reduce consumers' *perceived* privacy violations (PPV) is essential for evaluating their viability. In this research, we characterize advertising practices along the dimensions of tracking and targeting, and understand consumers' PPVs for current practices and proposed PETs through online experiments with U.S. and European consumers. As expected, the industry status quo of behavioral targeting, with high degrees of tracking and targeting, results in high PPV. While new technologies that keep data on users' devices reduce PPV compared to behavioral targeting, the reduction is minimal, including for group-level targeting. Contextual targeting, which involves no tracking, significantly lowers PPV. Not surprisingly, PPV is lowest when tracking is absent, but notably, consumers show similar preferences for untargeted ads and no ads. Importantly, consumer perceptions of privacy violations may not align with technical definitions, suggesting that operational investments in privacy technologies may fail without consumer validation. Therefore, it is essential for managers, industry practitioners, and policymakers to follow a consumer-centric approach to understanding privacy concerns and evaluating the operational viability of privacy-enhancing solutions.

## 1. Introduction

Firms use display advertising, including banner and video ads, to advertise to consumers online. With Internet penetration in 2025 estimated at approximately 98% in Northern Europe, 93% in Northern America, and 68% worldwide,[1] and over US $350 billion spent on display advertising worldwide,[2] a large majority of the world's population is exposed to display ads. Since the early 2000s, the dominant paradigm in display advertising has been behavioral targeting, under which an individual-level profile of a consumer is developed by tracking their Internet activity over time across websites and apps that they visit, and the consumer is targeted individually based on the profile.

Given such practices, consumer privacy concerns arise (Goldfarb and Tucker 2012; Martin et al. 2017). For example, Johnson (2013) observes that in the early 2010s, approximately two-thirds of Americans resisted behaviorally targeted advertising. More recent surveys suggest higher figures. For example, a survey by the Pew Research Center (2019) finds that 79% of Americans are concerned about how their data is collected and used by firms, and 81% feel that the potential risks of this data collection outweigh the benefits. Worledge and Bamford (2019) find that while a majority (63%) of individuals supported how digital advertising worked when initially asked, once a brief explanation of its functioning was provided, acceptability fell to 36%. Accountable Tech (2021) finds that 81% of Americans would keep their personal data private, even if it meant seeing less relevant ads, although other studies suggest that the framing of the question would suggest a higher acceptance rate (Prince and Wallsten 2022).

These privacy concerns impact the law and regulation of marketing practices, which can constrain or alter both the strategies and the operations of firms broadly (Staelin et al. 2023; Jaikumar et al. 2024) or specifically relating to privacy (Johnson et al. 2020; Jia et al. 2021; Marotta-Wurgler 2016). For example, privacy regulation such as the California Consumer Privacy Act (CCPA), the European Union's General Data Protection Regulation (GDPR), and the China Personal Information Protection Law (PIPL) may limit the opportunities of firms to collect and use user data (Jakhu and Chowdhury 2025). This concern is borne out by Google's abandonment of its Privacy Sandbox initiative in April 2025 (Reuters 2025) after almost six years of development. Google not only retreated from phasing out third-party cookies but subsequently announced the discontinuation of key Privacy Sandbox technologies, including Topics API and Protected Audience API[3]. While the push to remove cookies promised improved privacy, concerns regarding operational implementation and reduced utility of advertising ultimately led to the initiative's failure.

---

[1] https://www.statista.com/statistics/269329/penetration-rate-of-the-internet-by-region/.
[2] Global display ad spending was estimated to be greater than USD 352 billion in 2024: https://www.statista.com/statistics/276671/global-internet-advertising-expenditure-by-type/.
[3] https://privacysandbox.com/news/update-on-plans-for-privacy-sandbox-technologies/

From an operations management perspective, these privacy concerns present a fundamental challenge: how to design and operate advertising systems that balance commercial effectiveness with consumer privacy perceptions much like the distinction between perceived and objective quality (Dow et al. 1999; Schroeder et al. 2005). As Massimino et al. (2018) argue, digital confidentiality must be examined as an operations performance dimension alongside traditional metrics like cost, quality, and delivery. The advertising industry's response—developing privacy-enhancing technologies—requires massive operational investments in new infrastructure, technology adoption across multiple stakeholders, and fundamental changes to revenue management systems (Shen et al. 2021). Yet, without understanding how consumers actually perceive these technologies' privacy protections, firms risk wasteful capital expenditure on operationally complex solutions that fail to address the performance dimension they were designed to improve.

Consumer privacy concerns not only shape regulation and firm practices, but they, in turn, also influence how firms may create facts that shape consumer preferences and further regulation. Therefore, effective privacy regulation needs to appreciate the complexity of markets and, importantly, *consumer reactions and perceptions* (Hirshleifer 1980; Mayer and White 1969; Daviet et al. 2021). Experiments with consumers can inform how theoretical fundamentals related to the regulation of privacy and digital privacy strategies need to be revisited (Svirsky 2022). In this context, in this study, we conduct experiments to show how privacy-related implementations, even if well-intentioned, may diverge from how privacy protections from these implementations are *perceived* by consumers. The difference between privacy and perceived privacy is significant (Acquisti et al. 2013), and it has implications for marketing strategies such as targeted advertising using different technical approaches (Molitor et al. 2024) with different efficacies. Furthermore, industry practices on privacy protection that do not improve consumers' online experience and privacy perceptions would lead to wasteful expenditure in (re-)building the ad tech infrastructure yet not meeting their consumer privacy goals.[4]

Our research is in the context of *privacy-enhancing technologies (PETs)* (sometimes also referred to as Privacy-Enhancing Advertising Technologies (PEATs)), that have been developed by the online advertising industry to address consumers' privacy concerns and adapt to stricter global privacy laws. PETs are "digital technologies and approaches that permit the collection, processing, analysis, and sharing of information while protecting the confidentiality of personal data" (OECD 2023). PETs seek to preserve the utility of data while minimizing the necessity for extensive data collection and processing, thus preserving privacy. The most prominent examples of PETs in online advertising included initiatives under the Google Privacy Sandbox (Johnson 2024; Kobayashi et al. 2024; Gu et al. 2025), which Google discontinued in

[4] We note that our inquiry is related to "perceived quality" and "objective quality" that plays a significant role when addressing consumer perceptions (Mitra and Golder 2006).

2025.[5] These PETs were based on the idea of individuals' data not leaving their devices, while either targeting them in groups of consumers with shared interests (e.g., under the "Topics" approach), or enabling individual-level tracking, profiling, and ad serving (e.g., under the "Protected Audience" approach). We note that we focus exclusively on advertising-focused PETs and therefore we do not consider PETs like differential privacy and federated learning that are general-purpose privacy technologies. Having said that, we note that these underlying technologies may be used in developing advertising-focused PETs; for instance, both "Topics" and "Protected Audience" used federated learning.

PETs are distinct from behavioral targeting because they seek to preserve the utility of data and the ability to target effectively, while also preserving privacy (at least better than behavioral targeting),[6] and have been promoted by policymakers (Tucker 2023). While Google's PET efforts under the Privacy Sandbox have been unsuccessful, PETs themselves remain important. For example, the United States Office of Science and Technology Policy stated that PETs "present a key opportunity to harness the power of data and data analysis techniques in a secure, privacy-protecting manner."[7] The United States and United Kingdom governments have jointly announced a bilateral innovation prize to promote the advancement of PETs.[8] PETs have also been explicitly addressed in privacy and data protection laws and regulations worldwide (OECD 2023), such as in the European Union's GDPR, which states in Article 25 that PETs may help to implement the data protection principle of privacy by design and by default.[9] The advertising industry continues to develop and adopt PETs, with global investments in PETs projected to grow from approximately USD 2.4 billion in 2022 to USD 26 billion in 2029. Solutions other than Google's Privacy Sandbox being developed or deployed include Apple's Privacy Features (Intelligent Tracking Prevention (ITP), App Tracking Transparency (ATT), and Private Click Measurement), Mozilla Firefox's Anti-Tracking Tools (like Enhanced Tracking Protection (ETP)), Brave's Ad System, The Trade Desk's Unified ID 2.0, IAB Tech Lab's Project Rearc, Microsoft's PARAKEET, and W3C's Privacy Proposals.[10]

---

[5] https://privacysandbox.com/.

[6] Whether PETs actually improve consumer privacy is an open question. Edelman (2021), for example, argues that "Google's [Privacy Sandbox] is a classic example of what you might call privacy theater: While marketed as a step forward for consumer privacy, it does very little to change the underlying dynamics of an industry built on surveillance-based behavioral advertising."

[7] https://www.federalregister.gov/documents/2022/06/09/2022-12432/request-for-information-on-advancing-privacy-enhancing-technologies.

[8] https://www.whitehouse.gov/ostp/news-updates/2021/12/08/us-and-uk-to-partner-on-a-prize-challenges-to-advance-privacy-enhancing-technologies/.

[9] Specifically, the GDPR Article 25 (2) states that firms "shall implement appropriate technical and organizational measures for ensuring that, by default, only personal data which are necessary for each specific purpose of the processing are processed" (European Union 2016).

[10] See: https://www.kisacoresearch.com/content/investors-view-privacy-enhancing-technologies. Because Google is a dominant provider of online advertising services, Google's Privacy Sandbox solutions achieved high prominence during their development phase (Johnson (2025) finds that over 40% of the top 60,000 commercial websites on the Internet had adopted at least one of the Sandbox solutions by mid-2025). Following Google's announcement in April 2025 that it would not phase ot third-party cookies on Chrome, Google initially stated it would continue to

Evidently, the development of PETs has focused on the *technical and operational aspects* of how and where data tracking is done and how the data are used for targeting. While PETs have been developed in reaction to consumers' privacy concerns, curiously, an important aspect that has not been given due attention is *consumers' own perceptions* of the PETs being developed, such as how much consumers perceive that these practices violate or preserve consumer privacy. Presumably, PETs that vary in how they process and use data for tracking and targeting also vary in consumers' degrees of perception of how much their privacy is preserved or violated. Any disconnect between practices and perceived practices may create consumer and political backlash, which in turn may lead to changes in regulation and/or marketing practices.

In this research, we ask how different firm practices in online advertising impact consumers' *perceived privacy violation (PPV)*. Essentially, the question is how much consumers *perceive* their privacy to be violated or not when their data is being tracked in different ways and whether they are being shown targeted ads or not.[11] As operations management scholars increasingly recognize digital confidentiality as a critical performance dimension alongside traditional metrics (Massimino et al. 2018), understanding consumer privacy perceptions becomes essential for evaluating the operational viability of privacy-enhancing advertising technologies. We note that we do not measure consumers' stated willingness-to-pay (WTP) for privacy; our goal is only to measure their perceptions of privacy violations. Understanding these perceptions is critical because perceived privacy directly affects marketplace outcomes such as conversion rates and sales performance in digital platforms (Rong et al. 2022).

We measure consumers' PPV for various advertising practices, including two PETs, through an online study with several thousand consumers in the United States. The practices we investigate are behavioral targeting (in which tracking and targeting are at the individual-level, and the consumer's data leaves their machine), individual-level targeting PET (in which tracking and targeting are at the individual-level, and the consumer's data does not leave their machine), group-level targeting PET (in developing Privacy Sandbox solutions (https://privacysandbox.com/news/privacy-sandbox-next-steps/) and usage remained stable through June 2025 (https://www.linkedin.com/pulse/privacy-sandbox-one-play-garrett-johnson-kqdic/). However, Google has subsequently discontinued key Privacy Sandbox technologies including Topics API and Protected Audience API https://privacysandbox.com/news/update-on-plans-for-privacy-sandbox-technologies/, and the UK Competition and Markets Authority closed its Privacy Sandbox investigation. Our experimental study of these technologies remains relevant as it documents consumer perceptions of PET approaches that may resurface in future privacy initiatives and provides insights into why the Privacy Sandbox failed to achieve sufficient acceptance from consumers. Other privacy-preserving solutions continue to be developed are available at these links: https://www.apple.com/privacy/, https://www.mozilla.org/en-US/firefox/privacy/, https://brave.com/brave-ads/, https://www.thetradedesk.com/us/about-us/privacy/unified-id-2-0, https://iabtechlab.com/project-rearc/, https://learn.microsoft.com/en-us/microsoft-edge/privacy/web-privacy-whitepaper#parakeet, and https://www.w3.org/community/privacycg/.

[11] Spiekermann, Grossklags, and Berendt (2001) found that consumers "privacy concerns focused either on the revelation of identity aspects such as name, address or e-mail […] or on the profiling of interests, hobbies, health and other personal information […]."

which tracking is at the individual-level and targeting is at the group-level, and the consumer's data does not leave their machine), contextual targeting (in which only the consumer's presence on a specific website is used for targeting), and untargeted ads and no ads (in both of which there is no tracking or targeting).

We find that PPV for behavioral targeting is the highest. PPVs for the individual-level targeting PET and the group-level targeting PET are similar but slightly lower than those for behavioral targeting. This leads to the important learnings that: (i) PPV is reduced if data never leave a consumer's machine, but (ii) if past browsing data are used for profiling and targeting, the granularity at which they are used for targeting (i.e., group-level or individual-level) does not significantly impact PPV. Under contextual targeting, although there is individual targeting, the fact that there is no tracking across websites (i.e., no data collection and profiling) leads to relatively low PPV. Consumers' PPVs for seeing untargeted ads and no ads when they are not being tracked are similar and also the lowest among all practices.

The primary contribution of our research is that we contribute to understanding consumers' PPV for different current and proposed PETs in online advertising. We do not know of any other research that has studied PPV of different practices in this manner,[12] and believe that this research makes a significant contribution by helping to understand the consumer side of this critical question. We show that consumers perceive their privacy to be violated whenever their data are used to serve them ads—the violation felt is greater if their data leaves their machines, but is especially strong as the targeting goes from contextual to targeted based on browsing history (even if at the group-level). This understanding in itself has important implications for policymakers and advertisers. For instance, privacy regulations and industry efforts are focusing more on control of data and data security, rather than on the inferences that can be made with the data for targeting (Miklos-Thal et al. 2024). Our findings, however, suggest that consumers' perceptions of privacy are affected more by whether they will receive targeted ads and the experience they will have than by technological or operational aspects of how and where data are stored, how consumers are tracked, etc.[13] Hence, a consumer-centric approach to privacy may be warranted instead of relying solely on technical, engineering, or firm perspectives. We identify that marketing theory relating to regulation needs to better account for PPV, lest privacy policy by firms or regulators lead to unintended consequences that may hurt consumers and firm innovation. The learnings from our research may prompt a reevaluation of the current emphasis in privacy legislation and regulation as well as self-regulation.

[12] Lin (2022) develops a methodology to separate intrinsic and instrumental preferences of privacy for a specific practice but does not estimate the PPV of different practices. Prince and Wallsten (2022) elicit stated privacy preferences of consumers in different geographies and for different types of data and services. Tomaino, Wertenbroch, and Walters (2023) show that consumers have difficulties in stating their WTP for non-market goods (including privacy), and may give inconsistent answers even under incentive-aligned approaches. Relatedly, Strahilevitz and Kugler (2016) identify that consumers have a different understand of policies (2016).

[13] This notion is in line with Acquisti (2024) who states that Google Topics "can be privacy preserving, but it may not change how targeting ultimately operates in the online advertising ecosystem [...] that is, the fact that, even when their identities are nominally protected, individuals may be targeted with offers that may or may not be beneficial to them."

From an operations management standpoint, our research provides a framework for evaluating privacy technologies based on fundamental operational characteristics—tracking intensity and targeting granularity—that determine both technical complexity and consumer perceptions. This framework enables operations managers to assess whether proposed privacy technologies are likely to achieve sufficient consumer acceptance to justify operational restructuring costs. Our findings suggest that moderate privacy protections (like Privacy Sandbox's on-device processing) may fall into a "viability gap": operationally complex enough to require substantial infrastructure investment, yet insufficient to meaningfully reduce consumer privacy concerns. This has direct implications for technology adoption decisions, capacity planning for privacy infrastructure, and strategic choices about advertising operations architecture.

Second, as an additional contribution, we highlight a theoretical framework—specifically, dual-privacy theory (Lin 2022)—that is consistent with our experimental findings on when consumers perceive their privacy to be violated. Dual-privacy theory comprises of two components: (i) an intrinsic component and (ii) an instrumental component. The intrinsic component reflects a consumer's natural preference for privacy. The instrumental component, on the other hand, arises from a consumer's expected economic consequences of sharing their private information with the firm due to its use of this data. A consumer may perceive their privacy to be violated if one or both components of privacy preferences lead to disutility. Our findings show that prevailing PETs seem to address perceived violations about the intrinsic aspect of privacy, such as ensuring data remains on the consumer's device. However, these PETs may not effectively tackle the instrumental aspect of privacy, which involves how firms use consumers' data to target them, either at a group-level or at an individual-level. Therefore, the dual-privacy theory seems plausible in this context, which can be explored further in future research.

The rest of the paper is organized as follows. In the next section, we describe how we conceptualize our experiments and, following this, we present the results of our experiments. Having presented our experimental findings, we discuss how dual-privacy theory is consistent with them and has the potential to explain them. We conclude the paper with a discussion of the implications of our findings for advertisers and regulators.

## 2. Description of Experiments

In this section, we explain how we conceptualize and run our experiments.

### 2.1 Conceptualization of Experiments

In Table 1, we describe the main practices of firms in online advertising that we study, labeled as different scenarios from A to F. Each row corresponds to one scenario and describes it in two dimensions—tracking (how data are collected and whether data leave the user's machine or not) and targeting (how data are used, specifically, for targeting ads). It also associates a degree with each tracking and targeting practice with a higher degree indicating more involved or intense tracking and targeting. Note that while we consider six scenarios, among which are two PETs motivated by those that were proposed under Google Privacy Sandbox (prior to its discontinuation in 2025), the framework based on the dimensions of tracking and targeting is useful for understanding PPVs for other potential practices as well, because such practices can also be characterized along these dimensions.

For the status quo of behavioral targeting (Scenario F), the degree of tracking is high because the consumer is tracked across websites and the data leaves the local machine, and the degree of targeting is also high because the consumer is targeted at the individual-level. For the Individual-level Targeting PET (Scenario E), the degree of tracking is low because although the consumer's activity is tracked, the consumer's data does not leave the machine; however, the degree of targeting is high because the consumer receives individually-targeted ads. For the Group-level Targeting PET (Scenario D), the degree of tracking is low because although the consumer's activity is tracked, the consumer's data does not leave the machine. In this case, the degree of targeting is at a medium level because the consumer is profiled and receives ads targeted at a group-level.

Table 1:
Overview of Different Firm Practices in Online Advertising

| **Scenario** | **Online Advertising Practice** | **Tracking (Degree)** | **Targeting (Degree)** |
|---|---|---|---|
| A | No Ads, No Tracking | No tracking (Zero) | No targeting (Zero) |
| B | Untargeted Ads | No tracking (Zero) | No targeting (Zero) |
| C | Contextual Targeting | Individual-level determination of user presence on focal website but no past data used for profiling (Low) | Individual-level targeting based on context (Low) |
| D | Group-level Targeting PET (e.g., Google's discontinued "Topics") | Individual-level tracking but data stays on the user's machine (Low) | Group-level targeting based on behavior (Medium) |
| E | Individual-level Targeting PET (e.g., Google's discontinued "Protected Audience") | Individual-level tracking but data stays on the user's machine (Low) | Individual-level targeting based on behavior (High) |
| F | Behavioral Targeting | Individual-level tracking and data leaves machine (High) | Individual-level targeting based on behavior (High) |

For contextual targeting (Scenario C), the degrees of tracking and targeting are both low, as the focal website only determines the individual-level presence of the user on the website. The consumer is not tracked at the individual-level on the focal website they are visiting and on other websites (i.e., no past behavioral browsing data is used for profiling and targeting the user, and the only data used for targeting is the fact that the consumer is present on the website); however, contextual targeting may still trigger privacy concerns (Bleier 2021). When there is no tracking and no ads are shown to consumers (Scenario A), or untargeted ads are shown to consumers and they are not tracked (Scenario B), the degrees of tracking and targeting are both zero.

Postulating that a consumer's PPV is influenced by both the degree of tracking and the degree of targeting, based on the arguments presented, we expect consumers to have the highest PPV for Behavioral Targeting, followed by Individual-level Targeting PET, Group-level Targeting PET, Contextual Targeting, Untargeted Ads, and No Ads, in that order. Next, we report on the online experiments we ran to obtain data on the PPV of consumers in the United States.

### 2.2 Execution of Experiments

We conducted an online experiment in the United States to test consumers' PPV under various online advertising regimes. This experiment was guided by the arguments on the degrees of tracking and targeting in the last section. We report the details of our study below. In the Web Appendix, we report the results of two replication studies in the United States, as well as pooled results of our original U.S. study and the two U.S. replication studies. Finally, we report an additional replication study in Europe in the Web Appendix. The results of the additional studies are statistically identical to those of the original study presented below.

#### 2.2.1 Participants

We collected the data for our study through an online experiment on the platform Prolific on February 3, 2023. The study uses a survey to solicit consumers' PPV under the six experimental conditions representing various online advertising regimes. Stimuli and non-identifiable alphanumeric data will be available via an online data repository upon conditional acceptance of our paper. We prespecified when data collection would end (i.e., the decision to stop collecting data was independent of the results; we did not analyze the data until after data collection for the given study had been completed). Following recent thinking on sample size (www.datacolada.org/18), we sought to obtain a minimum of 250 participants per treatment group. Slight deviations from the target and actual numbers are caused by idiosyncratic differences in how survey "completes" are registered in Prolific vs. the survey software we used to collect the data. We report the results using all completed survey observations and remove incomplete observations.

### 2.2.2 Stimuli

We asked participants to read a short description of how online advertising could work in the future. We summarize the descriptions of the seven experimental conditions in Web Appendix A.1. Conditions A, B, D, E, and F correspond to Scenarios A, B, D, E, and F in Table 1. Conditions C1 and C2 correspond to Scenario C in Table 1 and are two variations of this scenario.[14]

### 2.2.3 Experimental procedure

We developed seven different independent experimental groups and used a between-subjects design. Each participant was randomly assigned to one treatment group. After reading a description of how online advertising could work in the future in that scenario, participants completed a survey. The survey included a measure of PPV, demographics, and other measures. To indicate their PPV, participants were asked to respond to the statement: "Based on the scenario described above, do you perceive your privacy to be violated?" on a scale of 1 (not at all) to 7 (very much so).[15]

### 2.2.4 Face Validity and Attention Check

We determine the face validity of our PPV measure by running the following analysis per treatment group. Across the respondents in the group, we correlate the elicited PPV with the consumer's tendency to delete cookies as answered by the question "How often do you delete your browser cookies?" measured on a scale from 1 (never) to daily (9). We interpret the measure of the frequency of cookie deletion as a proxy for a consumer's sensitivity to privacy.

In experimental conditions where privacy matters, we expect a positive correlation between consumers' sensitivity to privacy and their stated PPV. We indeed find positive and significant (though small) correlations in all conditions where consumers are told they will receive targeted advertising, independent of whether targeting refers to contextual or behavioral targeting (Scenario A (No Ads, No Tracking): $r$ = 0.024, $p$ = 0.510; Scenario B (Untargeted Ads): $r$ = 0.060, $p$ = 0.109; Scenario C (Contextual Targeting): 0.091, $p$ = 0.000; Scenario D (Group-level Targeting PET): 0.162, $p$ = 0.000;

[14] We note that our focus is on a stated-preference setting where consumers are asked to state how much they perceive their privacy to be violated under a given advertising practice. In revealed preference settings, where consumers actually make privacy choices, the evidence is mixed for whether consumers consider costs or benefits to be greater. Some papers show that consumers click and convert more on targeted ads (Goldfarb and Tucker 2011; Hoban et al. 2015). However, other papers show that when consumers are given the choice, a majority of them choose not to be tracked (Kollnig et al. 2022; Kesler 2023; Aridor et al. 2024; Cheyre et al. 2024; Fang 2024; Kraft et al. 2024).

[15] Our PPV measure is inspired by Kahneman, Knetsch, and Thaler (1986), who measure consumers' perceived fairness perceptions using a single-item stated preference survey. This approach has also been used more recently to study consumer perceptions (Friedman and Toubia 2022).

Scenario E (Individual-level Targeting PET): $r = 0.081$, $p = 0.025$; Scenario F (Behavioral Targeting): $r = 0.126$, $p = 0.001$). We conclude from this analysis that our PPV measure has face validity.

As an attention check (see Web Appendix A.9 for details), we compute the median response time for each experimental group. We exclude any observations whose response times were more than one-third faster than their group's median. We then compare this constrained sample to our complete sample and find no differences between the two samples. We conclude that our respondents passed the attention check. Moreover, response times do not differ statistically across experimental groups, either in the constrained or complete sample.

#### 2.2.5 Test-Retest Reliability

We determine the test-retest reliability of our PPV measure by replicating our original study twice using a U.S. sample. We conducted the first replication study on February 23, 2023 (20 days after the original study) and the second replication study on September 25, 2023 (almost eight months after the original study and after Google announced the general availability of the Privacy Sandbox for the web on September 7, 2023 to consumers[16]). Note that Google subsequently abandoned the Privacy Sandbox initiative in April 2025. The results of the two replication studies are statistically identical to the original study's results. We report the detailed results of our original U.S. study below and the results of our two U.S. replication studies in Web Appendix A.2. and A.3. In addition, we report the pooled results of all three U.S. studies in Web Appendix A.4. Finally, we conducted a third replication study on November 11, 2023, using a European Union sample.[17] We report the results in Web Appendix A.5 and do not find any significant differences between the results of the E.U. and the original U.S. studies. Figure A1 provides an overview of the timing of our four studies and the varying salience of information on the Google Privacy Sandbox as measured by Google Trends data for the keyword "Privacy Sandbox" in 2023. Despite varying and overall increasing information search on the Google Privacy Sandbox throughout 2023 our repeated measurements of PPV over time do not show evidence of significant consumer learning about PETs in this time frame.

## 3. Results

We compare PPV across the seven experimental groups in the above study. The summary statistics are reported in Table 2, and the histograms of responses are plotted in Figure 1. First, as expected, PPV is the lowest when there is no tracking (Conditions A and B) and is statistically the same irrespective of whether ads are not shown (Condition A) or shown (Condition B). Second, if a consumer is being tracked (Conditions C1, C2, D, E, and F), then PPV is statistically significantly higher than when a consumer is not being tracked. Third, among the conditions in which a consumer is tracked, PPV is lowest for contextual ads (Conditions C1 and C2), where tracking simply means detecting that the consumer is present at a specific website. Fourth, if there is individual-level tracking of activity (Conditions D, E, and F), PPV is statistically significantly higher than in Conditions C1 and C2. Among Conditions D, E, and F, PPV is lower and statistically the same for Conditions D and E, in which data does not leave the local machine. At the same time, the distinction between profiling and targeting at the group-level (Condition D) or individual-level (Condition E) does not matter for PPV. Finally, PPV is highest for Condition F, which corresponds to the status quo of behavioral targeting with individual-level tracking, profiling, and targeting with data leaving the local machine.[18]

[16] https://privacysandbox.com/news/privacy-sandbox-for-the-web-reaches-general-availability.
[17] Our EU study was targeted to the 27 EU member states, Austria, Belgium, Bulgaria, Croatia, Republic of Cyprus, Czech Republic, Denmark, Estonia, Finland, France, Germany, Greece, Hungary, Ireland, Italy, Latvia, Lithuania, Luxembourg, Malta, Netherlands, Poland, Portugal, Romania, Slovakia, Slovenia, Spain and Sweden.

In summary, our results show that if online advertisers are not tracking a consumer, PPV is low, and they are indifferent if they see ads or not. Increased tracking, targeting, and data leaving the machine contribute to a larger PPV. The proposal by the industry (for example, within the Google Privacy Sandbox) of developing PETs under which data never leaves a consumer's machine lowers consumers' PPV compared to the current industry status quo of behavioral targeting in which data leaves the consumer's machine. However, PPV of group-level targeting does not significantly differ from PPV of individual-level targeting. This may be because consumers may perceive that they are being targeted based on their interests anyway, even if at a group-level. The decrease in PPV from PETs, under which data does not leave the machine of the consumer, though statistically significantly different, is small relative to the current industry status quo of behavioral targeting. On the other hand, the decrease in PPV from contextual targeting is comparatively much larger.

[18] There is an eighth condition in the experiment, Condition G, with Tracking but No Ads. This condition is an unrealistic condition, but we include it for theoretical completeness. This is out of the scope of our theoretical conceptualization, and we do not have a prediction for consumers' PPV for this condition. In the online experiment, this condition had 250 subjects, a mean PPV of 5.924 with a SE of 0.094 and a CI of [5.739, 6.109]. This PPV is even higher than for the behavioral targeting scenario (Condition F). Potentially, this is because, in the context of our study, if consumers are tracked but not shown ads then they may be suspicious about what exactly is being done with their data.

Table 2:
PPV per Experimental Group (N = 1,751)

| Experimental Group | Experimental Group Description | N | Mean | SE | CI |
|---|---|---|---|---|---|
| A | No Ads, No Tracking | 265 | 1.864 | 0.096 | [1.676, 2.053] |
| B | Untargeted Ads | 239 | 2.096 | 0.105 | [1.889, 2.303] |
| C1 | Contextual Targeting A | 235 | 2.698 | 0.116 | [2.470, 2.926] |
| C2 | Contextual Targeting B | 246 | 2.748 | 0.111 | [2.530, 2.966] |
| D | Group-level Targeting PET | 275 | 4.465 | 0.105 | [4.258, 4.673] |
| E | Individual-level Targeting PET | 247 | 4.563 | 0.109 | [4.349, 4.777] |
| F | Behavioral Targeting | 244 | 5.221 | 0.107 | [5.011, 5.432] |

Figure 1:
PPV per Experimental Group (N = 1,751)

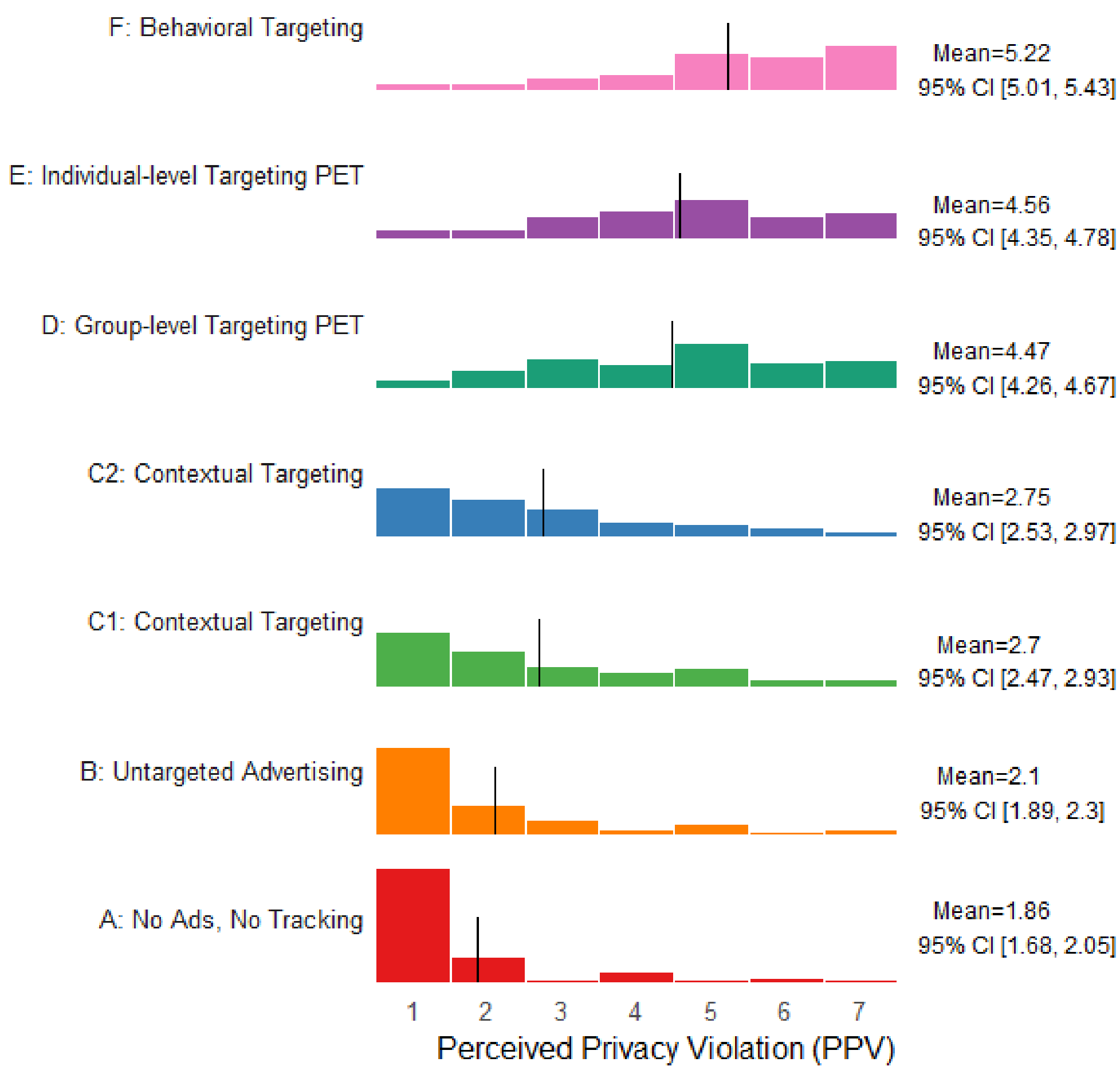

As some consumers may not have strong pre-existing privacy preferences and instead "construct" them in response to the context provided in our survey (see Dubé et al. 2025 for a more detailed discussion of the constructive preference view), we further investigate the role of context by testing whether providing additional information affects PPV. To do so, we vary the informational content under group-level targeting (Condition D) in a follow-up experiment which we conducted on July 31, 2024 (see Web Appendix A.7). We find that providing more information reduces PPV under Condition D2 relative to the original level of information provided under Condition D1; however, the relative rank order of contextual targeting, group-level targeting PET, and individual-level targeting PET remains unchanged.

Figure 2:
PPV per Experimental Group (N = 1,751)

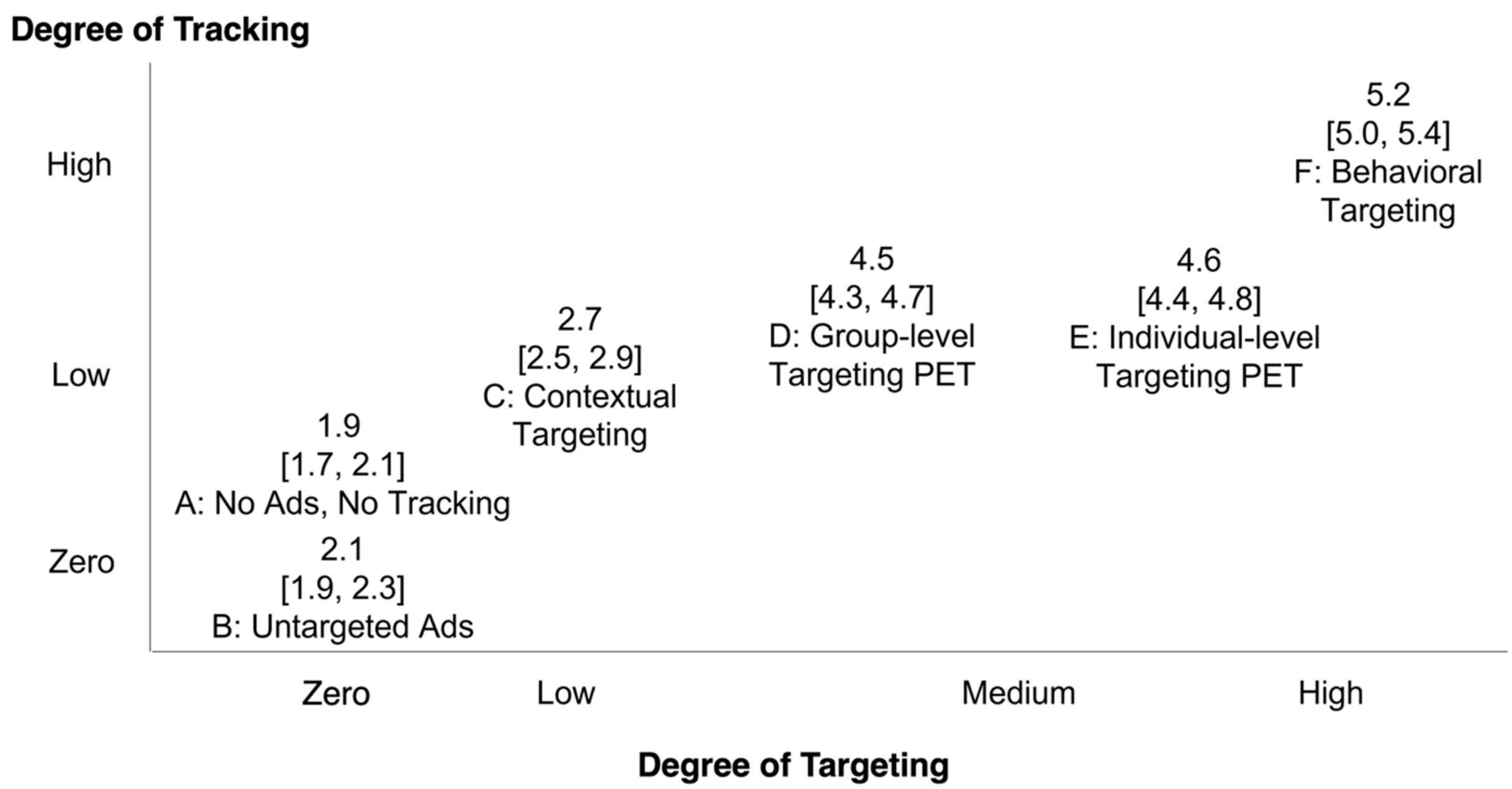

In Figure 2, we use the characterization of the different practices on the dimensions Degree of Tracking ($y$-axis) and Degree of Targeting ($x$-axis), and plot the results presented in Table 3 for the different conditions (for Conditions C1 and C2, we use the average PPV and plot it under Contextual Targeting). From eyeballing this figure, it is clear that as the degree of tracking, degree of targeting, or both increase for a particular practice (as per Table 1), the PPV for that practice weakly increases. This finding supports our key underlying assertion that the degrees of tracking and targeting are useful dimensions for conceptualizing and understanding different privacy-relevant practices related to tracking and targeting in online advertising, including new proposals such as under the Google Privacy Sandbox. This offers a valuable approach, rooted in fundamentals, to evaluate the privacy-related impact of firm practices in online advertising.

To further validate that consumers' PPV for different online advertising practices is based on both the degree of tracking and the degree of targeting, we conducted a follow-up experiment on July 31, 2024 with 1,499 participants in the United States (see Web Appendix A.8). In this experiment, participants made paired comparisons of the original experimental conditions, assessing their concerns on degrees of tracking and targeting. The findings demonstrate that participants do indeed invoke both, i.e., degrees of tracking and targeting, in their PPV evaluations and confirm the relative rankings of various advertising practices.

To further understand the relationship between the degrees of tracking and targeting and PPV, we run a descriptive regression using the pooled data from all three U.S. studies (with 5,193 total subjects) and report the results in Table 3. We use dummy-coding to include each expected level of degree of tracking (zero, low, high) and degree of targeting (zero, low, medium, high) per experimental group (see Table 1 for details) in the regression.[19]

As per Table 3, we find levels of degree of tracking (low: $\beta$ = 0.684, high: $\beta$ = 1.393) as well as levels of degree of targeting (medium: $\beta$ = 1.722, high: $\beta$ = 1.917) to be positively and highly statistically significantly (all $p$-values = 0.000) correlated with the consumers' PPV values. For higher (lower) levels of degrees of tracking and targeting, we find a stronger (weaker) positive relationship with PPV. These findings are in line with our arguments in Table 1. The Adjusted $R^2$ of the regression is 0.357.

Table 3:
Pooled Regression Results of Original Study and Two Replication Studies
for the Relationship of Degrees of Tracking and Targeting and PPV (N = 5,193)

| **Independent Variables** | | **Dependent Variable: Perceived Privacy Violation (PPV)** | **$p$-value** |
|---|---|---|---|
| Degree of Tracking | Zero | 0.000 | — |
| | Low | 0.684<br>(0.063) | 0.000 |
| | High | 1.393<br>(0.107) | 0.000 |
| Degree of Targeting | Low | 0.000 | — |
| | Medium | 1.722<br>(0.075) | 0.000 |
| | High | 1.917<br>(0.076) | 0.000 |
| Intercept | — | 1.967<br>(0.044) | 0.000 |

*Notes:* Standard errors in parentheses.

[19] "Zero" serves as baseline level for tracking, while "Low" serves as baseline level for targeting. Note that targeting = zero drops out of the estimation as it is perfectly colinear to tracking = zero.

## 4. Plausible Theory

After presenting our key findings, in this section we present a plausible theory consistent with our findings. We note that we simply present this as a theory with which our findings are consistent, and we do not rule out other possible theories.

We invoke dual-privacy theory, elements of which were initially proposed by Becker (1980), to explain how different firm practices in online advertising impact consumers' perceived privacy valuations (PPV). As further developed by Lin (2022), the dual-privacy theory consists of two components: intrinsic and instrumental.[20] The intrinsic component reflects a consumer's taste for privacy and arises from a desire to control one's personal information. The instrumental component reflects the economic consequences of revealing personal information; it includes the costs and benefits of sharing personal data with a firm through the firm's usage of these data.

According to Lin (2022), the intrinsic component is highly subjective and heterogenous across consumers and categories of data. Thus, some consumers may place a higher value on privacy than others and may be more likely to perceive privacy violations under different practices in online advertising, even if their personal data is not used in a harmful manner.

The instrumental component is also subjective, as consumers trade off the costs and benefits of sharing personal data with online advertising firms (Lin 2022). The costs may include consumers being charged higher prices or being shown too many ads once their type is known, and the benefits may include consumers being shown ads of products relevant to them. If the perceived costs of sharing personal data increase or the perceived benefits decrease, consumers are more likely to perceive their privacy to be violated.[21] In fact, consumers often attach a net negative perception to the instrumental aspects of privacy for behaviorally targeted ads, even if actual knowledge of behavioral advertising is low (Ur et al. 2012). For example, a 2012 Pew Research Center survey reported that 68% of the participants were "not okay with targeted advertising because [they do not] like having [their] online behavior traced and analyzed." In a more recent Pew Research Center survey (2019), 81% of the respondents stated, "that the potential risks they face because of data collection by companies outweigh the benefits." In a report for the European Commission, Armitage et al. (2023) note that the costs of behavioral targeting outweigh the benefits and call for a reform of the current online advertising business model. Lin et al. (2023) argue through ad blocking experiments that in general consumers perceive the welfare effects of ads to be negative. Mustri et al. (2024) argue through online experiments that if the counterfactual of what consumers would do if they are not shown ads is considered, then behaviorally targeted ads are unlikely to improve consumers' surplus.

[20] The ideas of intrinsic and instrumental components of privacy, sometimes along with this nomenclature, appear in various papers (Posner 1981; Calo 2011; Farrell 2012; Acquisti, Brandimarte, and Loewenstein 2015; Acquisti, Taylor, and Wagman 2016; Jin and Stivers 2017). However, Lin (2022) was the first to integrate these ideas into a holistic formal theory that can be neatly applied to studying privacy preferences. Choi, Jerath, and Sarvary (2023) apply these ideas to theoretical work on privacy.

[21] Beke et al. (2022) develop a measure that indicates the degree of acceptance of information collection by firms in different scenarios; this index is based on instrumental privacy tradeoffs (though Beke et al. (2022) do not use that terminology).

The dual-privacy theory can be applied to understand different firm practices in online advertising and develop implications of their impact on consumers' perceptions of privacy violations. In fact, there is a correspondence between our approach and dual-privacy theory. First, our tracking component, which is about whether and where data is collected, corresponds to the intrinsic aspect of privacy, which arises from a desire to control one's personal data and information. Second, our targeting component, which is about how the data are used for targeting, corresponds to the instrumental aspect of privacy, which arises from what inferences are made with the data and reflects the economic consequences of revealing personal information. Therefore, a higher degree of tracking and targeting corresponds to greater intrinsic and instrumental disutility, respectively. More specifically, the intrinsic disutility and instrumental disutility for different practices are the following: behavioral targeting: high, high; individual-level PET: low, high; group-level PET: low, medium; contextual targeting: low, low; untargeted ads and no ads without tracking: zero, zero.

To explore whether there is any support for our potential theory, we use the data from our second replication study (see further details below and Web Appendix A.3). This study is an identical replication of our original study with the sole difference that after a respondent stated their PPV, they were asked why they provided a specific PPV score.[22] We use these qualitative statements for textual analysis, specifically topic analysis, to further understand what our PPV measure captures. We use the popular topic modeling approach, LDA analysis, for our purposes.[23] We investigate whether the respondents' qualitative statements reflect intrinsic and instrumental privacy preferences.

As shown in the left panel of Figure 3, when we ask LDA to give two topics, we obtain one topic for which the highest-relevance keywords include "data," "privacy," "personal," "device(s)," and "leave," and another topic for which the highest-relevance keywords include "tracked/tracking," "ads/advertising," "activity," and "targeted." Based on these highest-relevance keywords, we can label the first topic as "Intrinsic Disutility" and the second topic as "Instrumental Disutility." As the right panel shows, the intrinsic and instrumental disutility topics account for 54.9% and 45.1% of the words in our corpus, respectively, and these topics are distinct (based on the inter-topic distance map). Overall, the topics we identify relate to intrinsic and instrumental disutility and provide additional support that the dual-privacy theory is a plausible theory to understand the PPVs of different advertising practices. However, we note that our theoretical exploration is indicative and not conclusive, and future work can study this further.

---

[22] The exact wording of the question is "Please explain why you stated a score of ["show previously stated PPV score"] for your perceived privacy violation based on the online advertising scenario described in the previous question?".

[23] Specifically, we use the Variational Expectation Maximization (VEM) algorithm (Blei et al. 2003). We used 9,444 words that appeared most frequently across the qualitative statements for the analysis. We exclude infrequent words (< 5 occurrences) to mitigate the risk of rare-word occurrences and co-occurrences confounding the topics. The remaining words used for analysis represent 68% of all words in the corpus. Based on our theoretical expectations motivated by the dual-privacy theory, we preset the number of topics for the LDA analysis to two.

Figure 3:
Two LDA Topics Representing Intrinsic Disutility and Instrumental Disutility

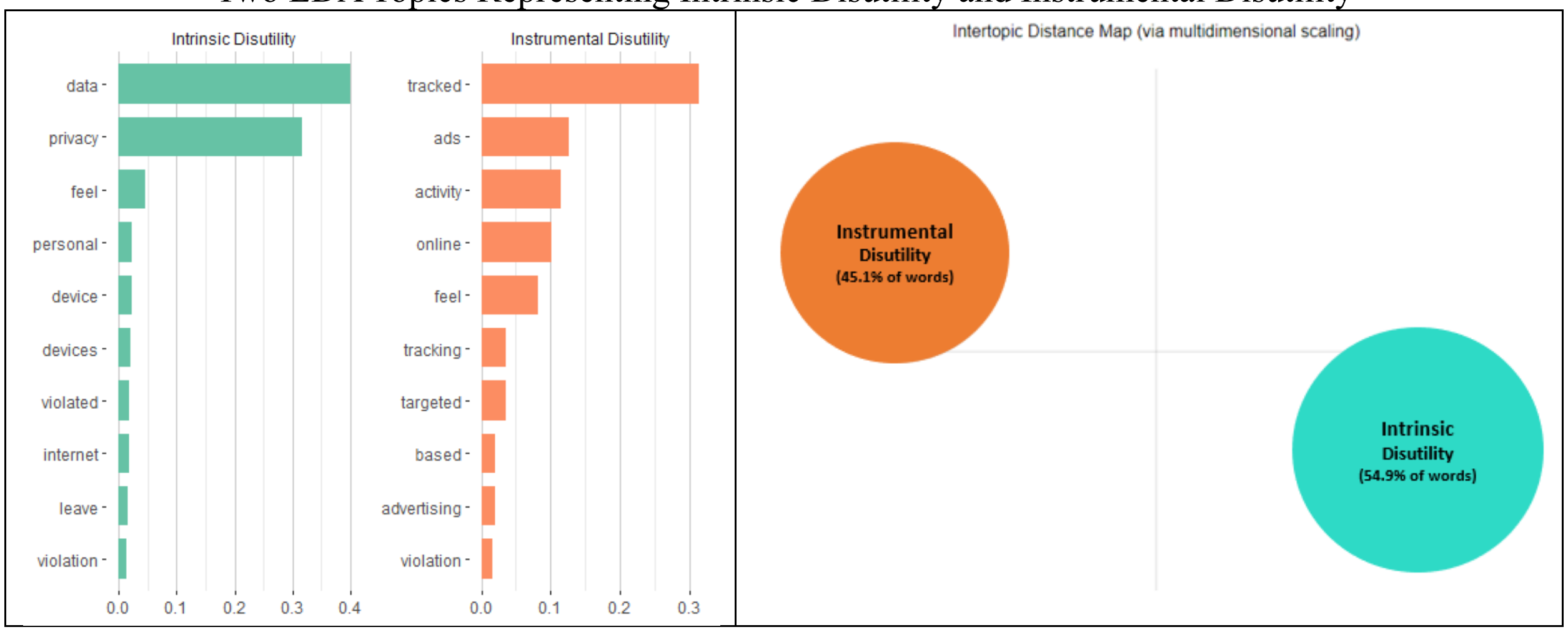

*Notes*: The words with the highest relevance for a topic are the words that have the highest probability to occur with a topic, i.e., the highest p(word|topic).

**5. Conclusions and Discussion**

This research examines consumers' perceived privacy violation (PPV) resulting from different firm practices in online advertising related to preserving consumers' privacy. We hypothesize that PPV depends on the degree of tracking and the degree of targeting under a practice. Using an online experiment with 1,751 US participants, we find that the current industry standard of behavioral targeting leads to a high PPV. We also investigate the PPV of privacy-enhancing technologies (PETs), such as group-level and individual-level targeting, with the data being kept on the consumer's machine. Our results show that while these PETs lower PPVs, the decrease is relatively small, and this small decrease is from the promise of data not leaving a consumer's machine reduces PPV rather than the promise of group-level targeting.

We contribute to our understanding of how consumers perceive their privacy to be violated for different privacy practices and proposals in online advertising, which has important implications for policymakers and advertisers. For instance, we find that something that conserves privacy from a technical point of view, such as group-level targeting, may not lead to a greater perception of privacy being preserved from a consumer's point of view if consumers perceive that targeting is still specific enough. We theorize that group-level targeting does not reduce PPV as much as perhaps anticipated because if users perceive that they are still being targeted at an individual-level, even if the targeting is group-based, this will prevent the privacy concerns from diminishing significantly. Group-level targeting was intended to improve the alignment of advertising strategy and operations, but end-users do not perceive those benefits. Thus, the promise of group-level targeting to improve privacy has not been realized.

Since the goal of privacy-enhancing initiatives is to cater to consumers' needs for privacy, firms and policymakers must take steps to enhance perceived and technical privacy to realize the gains of operational and marketing strategies. Such initiatives could include measures to change consumers' perceptions of not only the process of online advertising (i.e., consumers' understanding of the privacy-preserving nature of individual-level tracking without data leaving the local machine, such as under Google's discontinued Protected Audience) but also change the consumers' perceptions of the outcome of online advertising (i.e., being targeted with ads albeit on a more privacy-preserving group-level instead of the individual-level such as under Google's discontinued Topics). At the same time, consumer education on privacy initiatives may also be useful in bridging the gap between technical definitions of privacy and perceived privacy. The approach may need to be differentiated across large and smaller firms, for which regulators need to think through the potential consequences of such differentiation. Our empirical results on PPV align with predictions made under the assumption that the costs of sharing data for consumers are greater than the benefits that accrue to them; we interpret this as indirect support that, indeed, the general perception among consumers is that costs of sharing data are greater than the benefits (for which there is increasing evidence even in revealed-preference settings; see e.g., Kollnig et al. 2022; Kesler 2023; Aridor et al. 2024; Cheyre et al. 2024; Fang 2024; Kraft et al. 2024). For those who accept that there may be misalignment between PPV and actual privacy, these results are not surprising. However, these results support the need, often ignored in the policy community, that PPV is a real phenomenon. Indeed, many may not know that, under the various data policies, data is leaving their devices (Taylor 2004).

Current regulatory initiatives focus on the degree of tracking (control, collection, and data security). Our research shows that the degree of targeting (how the data are used for targeting, such as making inferences from it (Miklos-Thal et al. 2024)) deserve careful consideration as well.

Indeed, the PET solutions that were developed (like "Topics" and "Protected Audience" under the now-discontinued Google Privacy Sandbox initiative), which did not reduce PPV by much compared to behavioral targeting, did not mitigate targeting concerns, even though they mitigated tracking concerns. However, the practice of contextual targeting, which also mitigates targeting concerns, reduces PPV significantly compared to behavioral targeting.

The discontinuation of Google's Privacy Sandbox in 2025 offers important context for these findings. Google cited low adoption rates and ecosystem feedback indicating limited expected value as primary reasons for abandoning most Sandbox APIs, with the UK Competition & Markets Authority subsequently closing its oversight and releasing Google from its Privacy Sandbox commitments. Beyond these stated reasons, multiple adoption barriers contributed to the initiative's failure. The limited reduction in consumer PPV that we documented for Privacy Sandbox technologies provides insight into weak consumer demand for these solutions. Industry practitioners further highlighted that Privacy Sandbox was "built by engineers, for engineers"—technically over-complex and commercially misaligned[24]. Multiple ad tech firms participated in CMA-mandated testing but concluded there was no viable business case: implementation costs exceeded uncertain benefits for both advertisers and publishers. The operational complexity of online advertising revenue management—which requires sophisticated optimization of ad inventory allocation (Shen et al. 2021)—made Privacy Sandbox's disruptive approach particularly challenging for publishers to adopt. This confluence of factors—limited consumer privacy perception improvements, technical complexity, weak commercial outcomes, and high implementation costs—created insufficient market enthusiasm to overcome regulatory and competitive obstacles.

Our findings underscore a critical lesson for future privacy initiatives: successful PETs must simultaneously (1) substantially improve consumer privacy perceptions, (2) demonstrate clear commercial value, and (3) maintain reasonable implementation complexity. Technologies addressing only technical privacy (intrinsic aspects) without addressing how data are used for targeting (instrumental aspects) will struggle to gain market acceptance, regardless of their engineering sophistication.

Our findings have direct implications for managers in advertising platforms and publishing. Technology investment decisions should assess not only technical privacy protections but their impact on consumer privacy perceptions—technologies addressing only intrinsic privacy (data security, on-device processing) without addressing instrumental privacy (targeting practices) may fail to generate sufficient consumer demand to justify operational complexity and infrastructure investments. The Privacy Sandbox failure illustrates this risk: operationally complex systems deployed without validating consumer acceptance can result in wasted capital expenditure. Our tracking/targeting framework provides operations managers with a tool for preliminary assessment before committing to infrastructure investments, while our results suggest that contextual targeting—operationally simpler and more compatible with existing revenue management systems—delivers greater PPV reduction than complex PETs. Without substantial PPV improvement, privacy technologies will struggle to achieve the multi-stakeholder coordination necessary for successful implementation across advertisers, publishers, and technology providers.

[24] https://www.linkedin.com/posts/jochenschlosser_privacysandbox-adtech-digitaladvertising-activity-7386148928196608000-enM3/

Before concluding, we highlight that our research is only a first step in understanding PPV, and we show degrees of tracking and targeting as its antecedents. We argue that these correspond to intrinsic and instrumental preferences of privacy. Overall, our research provides an intriguing set of insights on PPV and highlights a plausible theory for understanding the PPV of different practices in online advertising. Future work can add to our findings and enrich them further.